\documentclass[prd,%
reprint, 
superscriptaddress,
showpacs,
nofootinbib,
 amsmath,amssymb,
 aps,
floats,
floatfix,
twocolumn]{revtex4-1}

\usepackage{graphicx}
\usepackage{dcolumn}
\usepackage{bm}
\usepackage{url}
\newcommand{\PR}[1]{\ensuremath{\left[#1\right]}} 
\newcommand{\PC}[1]{\ensuremath{\left(#1\right)}} 
\usepackage{hyperref}

\begin{document}


\title{Dynamics of Linear Perturbations in the hybrid metric-Palatini gravity}

\author{Nelson A. Lima}
 \email{ndal@roe.ac.uk}
\affiliation{Institute for Astronomy, University of Edinburgh, Royal Observatory, Blackford Hill, Edinburgh, EH9 3HJ, UK}

\renewcommand{\abstractname}{Abstract}
\begin{abstract}
In this work we focus on the evolution of the linear perturbations in the novel hybrid metric-Palatini theory achieved by adding a $f(\mathcal{R})$ function to the gravitational action. Working in the Jordan frame, we derive the full set of linearized evolution equations for the perturbed potentials and present them in the Newtonian and synchronous gauges. We also derive the Poisson equation, and perform the evolution of the lensing potential, $\Phi_{+}$, for a model with a background evolution indistinguishable from $\Lambda$CDM. In order to do so, we introduce a designer approach that allows to retrieve a family of functions $f(\mathcal{R})$ for which the effective equation of state is exactly $w_{\textrm{eff}} = -1$. We conclude, for this particular model, that the main deviations from standard General Relativity and the Cosmological Constant model arise in the distant past, with an oscillatory signature in the ratio between the Newtonian potentials, $\Phi$ and $\Psi$.
\end{abstract}

\pacs{98.80.-k, 95.36.+x, 04.50.Kd \hfill \today}
\maketitle


\section{\label{Int}Introduction}

Einstein's General Relativity (GR) is modern cosmology's main framework, providing a set of equations that dictate the dynamics of our Universe according to its material constituents. By them, our Universe could be expanding, static, or even collapsing. However, it is now well established that our Universe is currently undergoing an accelerated expansion which was preceded by phases of matter and radiation domination where gravitational attraction resulted in a decelerated expansion. And, at the beginning, it should have experienced a period of quasi-exponential inflation, so that any primordial spatial curvature would have been wiped out, leading to the spatially-flat and homogeneous Universe we observe.

The simplest explanation for the Universe's accelerated expansion is a cosmological constant, $\Lambda$, with a constant ratio of pressure to density (usually defined as the equation of state, $w$) equal to $-1$. Despite being in agreement with supernovae observations \cite{accel1,accel2,accel3,accel4}, data from the cosmic microwave background (CMB) \cite{cmb1,cmb2} including the recent {\it Planck} data \cite{planck1}, and large-scale structure (LSS) data \cite{lss1}, cosmologists still struggle to account for the difference between the theoretically-expected value for its energy density and the observed one. If it exists, observationally, it should account for approximately $70 \%$ of the Universe's total energy density, a value $121$ orders of magnitude smaller than that obtained from quantum field theory (for a review on $\Lambda$, see Ref.~\cite{lambdareview}). 

In light of these issues, new physics may be in order to account for that major component of our Universe, usually labeled dark energy (DE). Some theories, such as quintessence, k-essence, and so on, propose scalar fields rolling in a potential (see Ref.~\cite{quintessence} and references therein for a comprehensive review). Other theories consider higher dimensions, as in braneworld models such as the DGP model \cite{DGP,branes}, or assume that GR fails on cosmological scales and propose corrections to Einstein's action. The latter are grouped as the so-called Modified Gravity Theories (MGT), such as the Brans--Dicke scalar--tensor theory \cite{bd}, Galileon models \cite{gall}, the Fab Four \cite{fabfour}, $f(R)$ theories \cite{frreview}, and many others. For an extensive review on MGT, see Ref.~\cite{reviewall}.

In this paper, we focus on  a novel model, the hybrid metric-Palatini gravity \cite{main1,main2}. In this type of theories, the usual Einstein-Hilbert Lagrangian is supplemented with an $f(\mathcal{R})$ Palatini term. This type of hybrid theory arises when perturbative quantization methods are considered on Palatini backgrounds \cite{quantization}, which are connected with non-perturbative quantum geometries \cite{quantumgeo}.

Like the pure metric and Palatini cases, the hybrid theory has a dynamically equivalent scalar-tensor representation \cite{main1,main2}. Those authors have also shown that the scalar field need not be massive in order to pass the stringent Solar System constraints \cite{main1}, in contrast to the metric $f(R)$ theories, while possibly modifying the cosmological \cite{hybridcosmo} and Galactic \cite{hybridgala} dynamics due to its light, long-range interacting nature. In Ref. \cite{hybridcosmo}, the criteria for obtaining cosmic acceleration was discussed. Alongside that, several cosmological solutions were derived, depending on the form of the effective scalar field potential, describing both accelerating and decelerating Universes.

In this work, we focus on cosmological perturbations in the hybrid metric-Palatini formalism in the Jordan frame. Therefore, in Section \ref{I} we briefly introduce the hybrid metric-Palatini model and, in Section \ref{designer}, we present the designer approach which allows us to retrieve a family of solutions for $f(\mathcal{R})$ whose effective equation of state is $w_{\textrm{eff}} = -1$. In Section \ref{II} we derive the full set of perturbed cosmological equations and present them in the Newtonian and Synchronous gauges. Then, in Section \ref{III} we derive the Poisson equation and re-express the perturbed potentials in terms of the lensing potential, $\Phi_{+}$, and the slip, $\chi$, which we numerically evolve. We finish in Section \ref{conclusion} with the conclusions of this work.

\section{\label{I}Description of the hybrid metric-Palatini gravity}

The four-dimensional action describing the hybrid metric-Palatini gravity is given by
\begin{equation}{\label{action}}
 S = \frac{1}{2\kappa^2} \int d^4 x \sqrt{-g}\PR{R + f(\mathcal{R})} + S_{m},
\end{equation}
\noindent where $\kappa^2 = 8 \pi G$ and we set $c=1$. $S_m$ is the standard matter action, $R$ is the metric Einstein-Hilbert Ricci scalar and $\mathcal{R} = g^{\mu \nu} \mathcal{R}_{\mu \nu}$ is the Palatini curvature. The latter is defined in terms of the metric elements, $g^{\mu \nu}$, and a torsion-less independent connection, $\hat{\Gamma}$, through
\begin{equation}
\mathcal{R} \equiv g^{\mu \nu}\PC{\hat{\Gamma}^{\alpha}_{\mu \nu, \alpha} - \hat{\Gamma}^{\alpha}_{\mu \alpha,\nu} + \hat{\Gamma}^{\alpha}_{\alpha \lambda} \hat{\Gamma}^{\lambda}_{\mu \nu} - \hat{\Gamma}^{\alpha}_{\mu \lambda} \hat{\Gamma}^{\lambda}_{\alpha \nu}}.
\end{equation}

Varying the action (\ref{action}) with respect to the metric, one obtains the usual set of Einstein equations, given by
\begin{equation}{\label{einstein}}
 G_{\mu \nu} + F(\mathcal{R})\mathcal{R}_{\mu \nu} - \frac{1}{2}f(\mathcal{R})g_{\mu \nu} = \kappa^2 T_{\mu \nu},
\end{equation}
\noindent where $G_{\mu \nu}$ is Einstein's tensor, and $F(\mathcal{R}) \equiv df(\mathcal{R})/d\mathcal{R}$. $T_{\mu \nu}$ is the matter field's stress--energy tensor, defined as
\begin{equation}
 T_{\mu \nu} = -\frac{2}{\sqrt{-g}} \frac{\delta\PC{\sqrt{-g} \mathcal{L}_m}}{\delta \PC{g^{\mu \nu}}},
\end{equation}
\noindent where $\mathcal{L}_{m} \equiv \mathcal{L}_{m} \PR{\chi_i, g_{\mu \nu}}$ is the minimally--coupled matter Lagrangian, dependent on the matter fields, $\chi_i$, and the metric elements, $g_{\mu \nu}$.

On the other hand, varying the action (\ref{action}) with respect to the independent connection, $\hat{\Gamma}^{\alpha}_{\mu \nu}$, one gets the following equation,
\begin{equation}{\label{independent}}
 \hat{\nabla}_{\alpha}\PC{\sqrt{-g}F(\mathcal{R})g^{\mu \nu}} = 0,
\end{equation}
\noindent which implies that $\hat{\Gamma}^{\alpha}_{\mu \nu}$ is the Levi--Civita connection of a metric $h_{\mu \nu} = F(\mathcal{R})g_{\mu \nu}$. Hence, one can verify that the Palatini Ricci tensor, $\mathcal{R}_{\mu \nu}$, is related to the metric one, $R_{\mu \nu}$, by the following relation
\begin{eqnarray}{\label{riccitrelation}}
 \mathcal{R}_{\mu \nu} &=& R_{\mu \nu} + \frac{3}{2} \frac{1}{F^{2}(\mathcal{R})}F(\mathcal{R})_{,\mu}F(\mathcal{R})_{,\nu} \nonumber \\
 &&- \frac{1}{F(\mathcal{R})}\nabla_{\mu}F(\mathcal{R})_{,\nu} - \frac{1}{2}\frac{1}{F(\mathcal{R})}g_{\mu \nu} \Box F(\mathcal{R}).
\end{eqnarray}

Introducing an auxiliary scalar field, $\phi$, the action (\ref{action}) can be recast into a scalar-tensor theory described by the following action \cite{main1}
\begin{equation}{\label{scalaraction}}
 S = \frac{1}{2 \kappa^2} \int d^4 x \sqrt{-g} \PR{R + \phi \mathcal{R} - V(\phi)} + S_{m},
\end{equation}
\noindent where $\phi \equiv F(\mathcal{R})$ and $V(\phi) = \mathcal{R} F(\mathcal{R}) - f(\mathcal{R})$. Varying the latter action with respect to the metric $g_{\mu \nu}$, the scalar $\phi$, and the independent connection $\hat{\Gamma}^{\alpha}_{\mu \nu}$, leads to
\begin{equation}{\label{einsteinscalar}}
 R_{\mu \nu} + \phi \mathcal{R}_{\mu \nu} - \frac{1}{2}\PC{R + \phi \mathcal{R} - V}g_{\mu \nu} = \kappa^2 T_{\mu \nu},
\end{equation}
\begin{equation}{\label{palariccis}}
 \mathcal{R} - V_{\phi} = 0,
\end{equation}
\begin{equation}
 \hat{\nabla}_{\alpha} \PC{\sqrt{-g} \phi g^{\mu \nu}} = 0,
\end{equation}
\noindent respectively. Equation (\ref{riccitrelation}) can be easily rewritten in terms of the auxiliary scalar field by using $F(\mathcal{R}) = \phi$ which, when traced, leads to the following relation between the metric and the Palatini Ricci scalars:
\begin{equation}{\label{riccisrelation}}
 \mathcal{R} = R + \frac{3}{2\phi^2}\PC{\partial \phi}^{2} - \frac{3}{\phi} \Box \phi.
\end{equation}

Using Eqs. (\ref{riccitrelation}) and (\ref{riccisrelation}), one can rewrite Eq. (\ref{einsteinscalar}) as
\begin{eqnarray}{\label{einsteinscalar2}}
 G_{\mu \nu}\PC{1+\phi} &=& \kappa^2 T_{\mu \nu} - \frac{3}{2 \phi}\partial_{\mu} \phi \partial_{\nu} \phi + \nabla_{\mu} \nabla_{\nu} \phi \nonumber \\
 &&- \frac{g_{\mu \nu}}{2}\PC{V + 2 \Box \phi} + \frac{3}{4 \phi} g_{\mu \nu} \PC{\partial \phi}^{2}. 
\end{eqnarray}
\noindent The latter equation can be recast in the usual GR form $G_{\mu \nu} = \kappa^{2}T_{\mu \nu}^{\textrm{hybrid}}$, such that
\begin{eqnarray}
 \PC{1+\phi}T_{\mu \nu}^{\textrm{hybrid}} &=& T_{\mu \nu} + \frac{1}{\kappa^2} \Bigg[ \nabla_{\mu} \nabla_{\nu} \phi - \frac{3}{2 \phi}\partial_{\mu} \phi \partial_{\nu} \phi \nonumber \\
 &-& \frac{g_{\mu \nu}}{2}\PC{V + 2 \Box \phi} + \frac{3}{4 \phi} g_{\mu \nu} \PC{\partial \phi}^{2} \Bigg]
\end{eqnarray}
\noindent Adopting a flat Friedmann-Robertson-Walker (FRW) metric, $ds^2 = - dt^2 + a^{2}(t)d\vec{x}^2$, one can get the modified Friedmann equations:
\begin{equation}{\label{hubble}}
 3H^{2} = \frac{1}{1 + \phi} \PR{\kappa^{2} \rho - 3 H \dot{\phi} - \frac{3 \dot{\phi}^{2}}{4 \phi} + \frac{V(\phi)}{2}},
\end{equation}
\noindent and,
\begin{equation}
 2\dot{H} = \frac{1}{1 + \phi} \PR{ -\kappa^{2}\PC{\rho + p} + H \dot{\phi} - \ddot{\phi} + \frac{3 \dot{\phi}^{2}}{2 \phi}},
\end{equation}
\noindent where a dot stands for a differentiation with respect to time, $t$, and $H = \dot{a}/a$ is the Hubble parameter.

Lastly, closing the set of cosmological equations, one can trace Eq. (\ref{einsteinscalar}), getting
\begin{equation}{\label{tracescalar}}
 \frac{\phi}{3}\PR{2V - \PC{1 + \phi}\frac{dV}{d\phi}} - \Box \phi + \frac{1}{2\phi}\PC{\partial \phi}^{2} = \frac{\kappa^2 \phi}{3} T,
\end{equation}
\noindent where we have used Eqs. (\ref{palariccis}) and (\ref{riccisrelation}), and $T$ is the stress--energy tensor trace. Using the FRW metric, this equation takes the form
\begin{equation}
 \ddot{\phi} + 3H\dot{\phi} - \frac{\dot{\phi}^{2}}{2\phi} + \frac{\phi}{3}\PR{2V - \PC{1+\phi}\frac{dV}{d\phi}} = \frac{\kappa^2 \phi}{3}T,
\end{equation}
\noindent describing the dynamical evolution of the additional degree of freedom introduced by this theory.

Solving Eq. (\ref{tracescalar}) in the weak--field limit and far from matter sources, it has been shown that the scalar field behaves like \cite{main1,main2,hybridgala}
\begin{equation}
 \phi(r) \approx \phi_0 + \frac{2G\phi_0 M}{3r}e^{-m_{\phi}r},
\end{equation}
\noindent where the field's effective mass is defined as
\begin{equation}
 m^{2}_{\phi \vert \phi = \phi_{0}} = \frac{2V - V_{\phi} - \phi \PC{1 + \phi} V_{\phi \phi}}{3},
\end{equation}
\noindent where $\phi_0$ is the field's background value and $M$ is the distant source's mass, determined assuming spherical symmetry. The authors of Refs. \cite{main1,main2} have also obtained the solutions to the metric perturbations, and defined an effective Newton constant, $G_{\textrm{eff}}$, and the post-Newtonian parameter, $\gamma$, as, respectively
\begin{eqnarray}
  \label{effectiveG}
  G_{\textrm{eff}} &\equiv& \frac{G}{1+\phi_0}\PR{1 - \frac{\phi_0}{3}e^{-m_\phi r}},\\
  \label{gammapara}
  \gamma &\equiv& \frac{1 + \phi_{0}/3 e^{-m_\phi r}}{1 - \phi_{0}/3 e^{-m_\phi r}}.
\end{eqnarray}

From Eqs. (\ref{effectiveG}) and (\ref{gammapara}), it becomes clear that $G_{\textrm{eff}} \approx G$ and $\gamma \approx 1$ as long as $\phi_0$ is small, regardless of the value of $m^{2}_{\phi}$. Hence, contrary to what is seen in metric $f(R)$ theories, the hybrid metric-Palatini theory does not seem to need an evading mechanism, such as the chameleon mechanism, to pass the Solar System tests.

\section{$\Lambda$CDM Designer Approach}{\label{designer}}

In this section, we present a way to numerically get a family of $f(\mathcal{R})$ functions that reproduce a $\Lambda$CDM-like background evolution. Starting from Eq. (\ref{einstein}), it is clear one can define an effective stress--energy tensor for the hybrid metric-Palatini theory as
\begin{equation}{\label{hybridT}}
 T_{\mu \nu}^{\textrm{hybrid}} = T_{\mu \nu} + T_{\mu \nu}^{\textrm{eff}},
\end{equation}
\noindent where $T_{\mu \nu}^{\textrm{eff}}$ will be given by
\begin{equation}{\label{effT}}
 T_{\mu \nu}^{\textrm{eff}} = \frac{1}{\kappa^2}\PR{f\PC{\mathcal{R}}\frac{g_{\mu \nu}}{2} - F\mathcal{R}_{\mu \nu}},
\end{equation}
\noindent where we have omitted the dependence on $\mathcal{R}$ of $F$.

Using the relation between $R_{\mu \nu}$ and $\mathcal{R}_{\mu \nu}$ given by Eq. (\ref{riccitrelation}), one can define an effective equation of state
\begin{equation}{\label{weff}}
 w_{\textrm{eff}} = \frac{f \PC{\mathcal{R}} - 2F\PC{\dot{H} + 3H^2} - 5H\dot{F} - \ddot{F}}{-f\PC{\mathcal{R}} + 6F\PC{\dot{H} + H^2} - 3\dot{F}^{2}/F + 3H\dot{F} + 3\ddot{F}}.
\end{equation}
\noindent Taking $w_{\textrm{eff}} = -1$ in order to get a background evolution that follows $\Lambda$CDM, one gets a nonlinear second--order differential equation for $F$
\begin{equation}{\label{Fevolutiont}}
 \ddot{F} - H\dot{F} + 2 F \dot{H} - \frac{3}{2}\frac{\dot{F}^{2}}{F} = 0. 
\end{equation}
\noindent Changing to a logarithmic variable, i.e., $dt \rightarrow d \ln a$, such that $d/dt = H d/ d\ln a$, one gets
\begin{equation}{\label{F}}
 F^{\prime \prime} + F^{\prime}\PC{\frac{E^{\prime}}{2E} - 1} + F \frac{E^{\prime}}{E} - \frac{3}{2}\frac{F^{\prime 2}}{F} = 0,
\end{equation}
\noindent where we have define $E\PC{a} \equiv H^2/H_{0}^{2} = \Omega_{\rm{m}}a^{-3} + \Omega_{\gamma}a^{-4} + \Omega_{\textrm{eff}}a^{3 \int_{a}^{1} \PC{1+w_\textrm{eff}}d\ln a}$. In a flat Universe, $\Omega_{\textrm{eff}} = 1 - \Omega_{\rm{m}} - \Omega_{\gamma}$ and, for $w_{\textrm{eff}} = -1$, one recovers a $\Lambda$CDM-like cosmology.

In order to set the initial conditions, we consider an initial redshift, $z_{i} \gg 0$, but after the matter--radiation equality epoch. In these circumstances, the contribution from the effective component is negligible, meaning $E \approx \Omega_{\rm{m}} a_{i}^{-3} + \Omega_{\gamma} a_{i}^{-4}$. Hence, $E^{\prime}/E$ is reduced to a constant, given by
\begin{equation}{\label{ratioEprimeE}}
 \frac{E^{\prime}}{E} = - \frac{3 + 4 r_i}{1 + r_i},
\end{equation}
\noindent where $r_{i} = a_{\textrm{eq}}/a_{i}$ and $a_{\textrm{eq}} = \Omega_{\gamma}/\Omega_{\rm{m}}$. Therefore, at $z_{i}$, we have the following differential equation
\begin{equation}{\label{initialdifeq}}
 F^{\prime \prime} - F^{\prime}\PR{\frac{5 + 6 r_i}{2\PC{1+r_i}}} - F \PR{\frac{3 + 4 r_i}{1 + r_i}} - \frac{3}{2}\frac{F^{\prime 2}}{F} = 0,
\end{equation}
\noindent which happens to have a simple solution for the initial value of $F$, given by
\begin{equation}{\label{initialF}}
 F_{i} = C_{1} a_{i}^{-a} \PR{\cos \PC{\frac{1}{2}\PR{ \ln a_i + C_{2}} \sqrt{2 b - a^{2}}}}^{-2},
\end{equation}
\noindent where $a$ and $b$ are the coefficients multiplying $F^{\prime}$ and $F$ on equation (\ref{initialdifeq}), respectively, and $C_1$ and $C_2$ are constants that define the family of $f\PC{\mathcal{R}}$ functions yielding a $\Lambda$CDM-like evolution. According to our assumptions, $ 0 \leq r_i \leq 1$, meaning that $2 b - a^2$ will always be negative. Therefore, setting $2b - a^2 = - d$, where $d > 0$, our solution will, in fact, depend on a hyperbolic function,
\begin{equation}{\label{initialF2}}
 F_{i} = C_{1} a_{i}^{-a} \PR{\cosh \PC{\frac{1}{2}\PR{ \ln a_i + C_{2}} \sqrt{d}}}^{-2}.
\end{equation}
\noindent Finally, differentiating with respect to $\ln a_i$, one gets the initial condition for $F^{\prime}$
\begin{equation}{\label{initialFprime}}
 F^{\prime}_{i} = - C_{1} \frac{a_{i}^{-a}}{\cosh \PC{...}^{2}} \PR{a + \sqrt{d} \tanh \PC{...}},
\end{equation}
\noindent where we omitted the argument for the hyperbolic functions, defined in Eq. (\ref{initialF2}). Since $a > \sqrt{d}$, from Eq. (\ref{initialFprime}) one can see that $F^{\prime}_{i}$ cannot be set to zero. Hence, we should expect some deviation from standard General Relativity at the beginning of evolution.

\begin{figure}[t!]
\begin{center}$
\begin{array}{c}
\includegraphics[scale = 0.44]{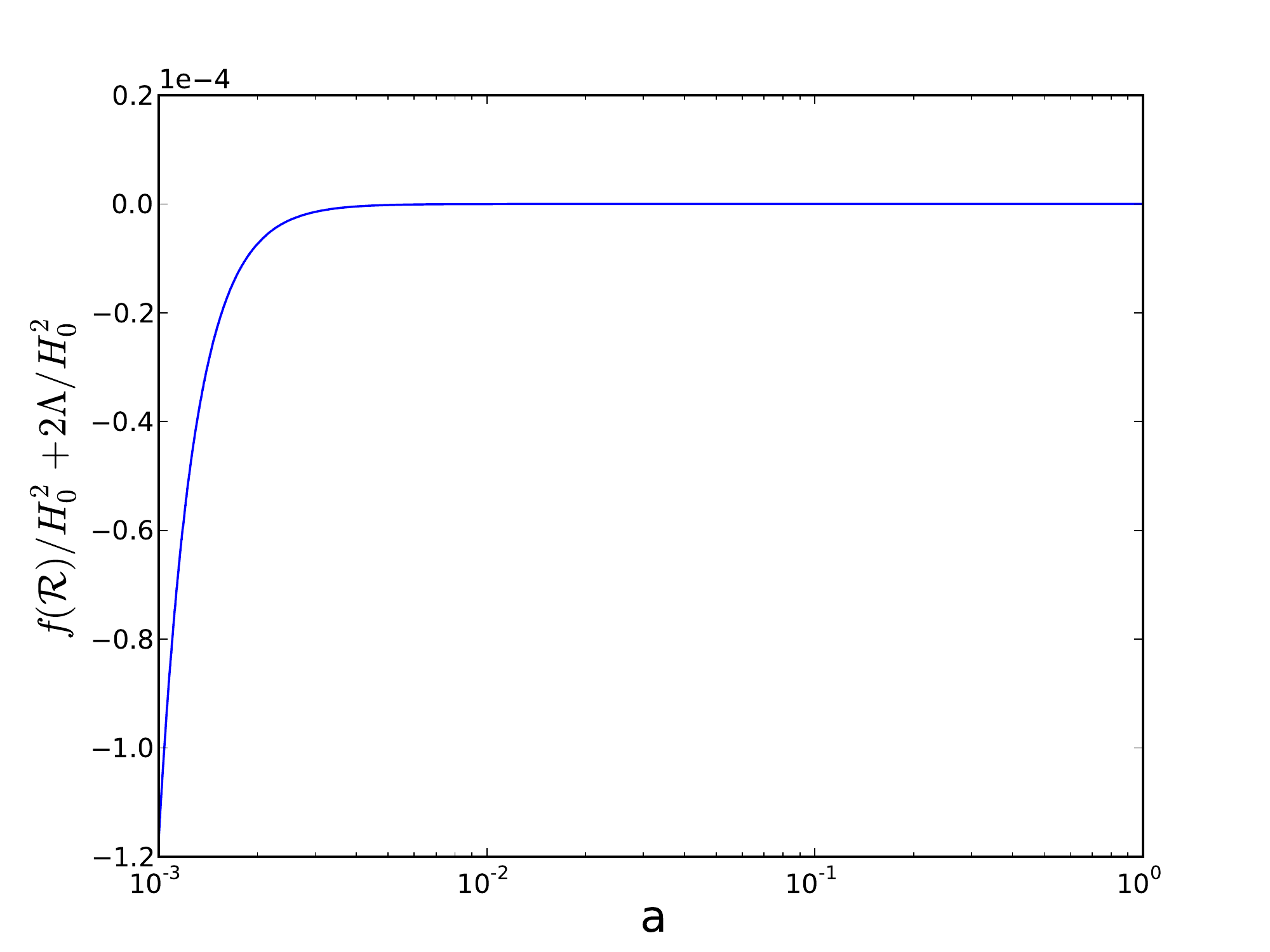} \\
\includegraphics[scale = 0.44]{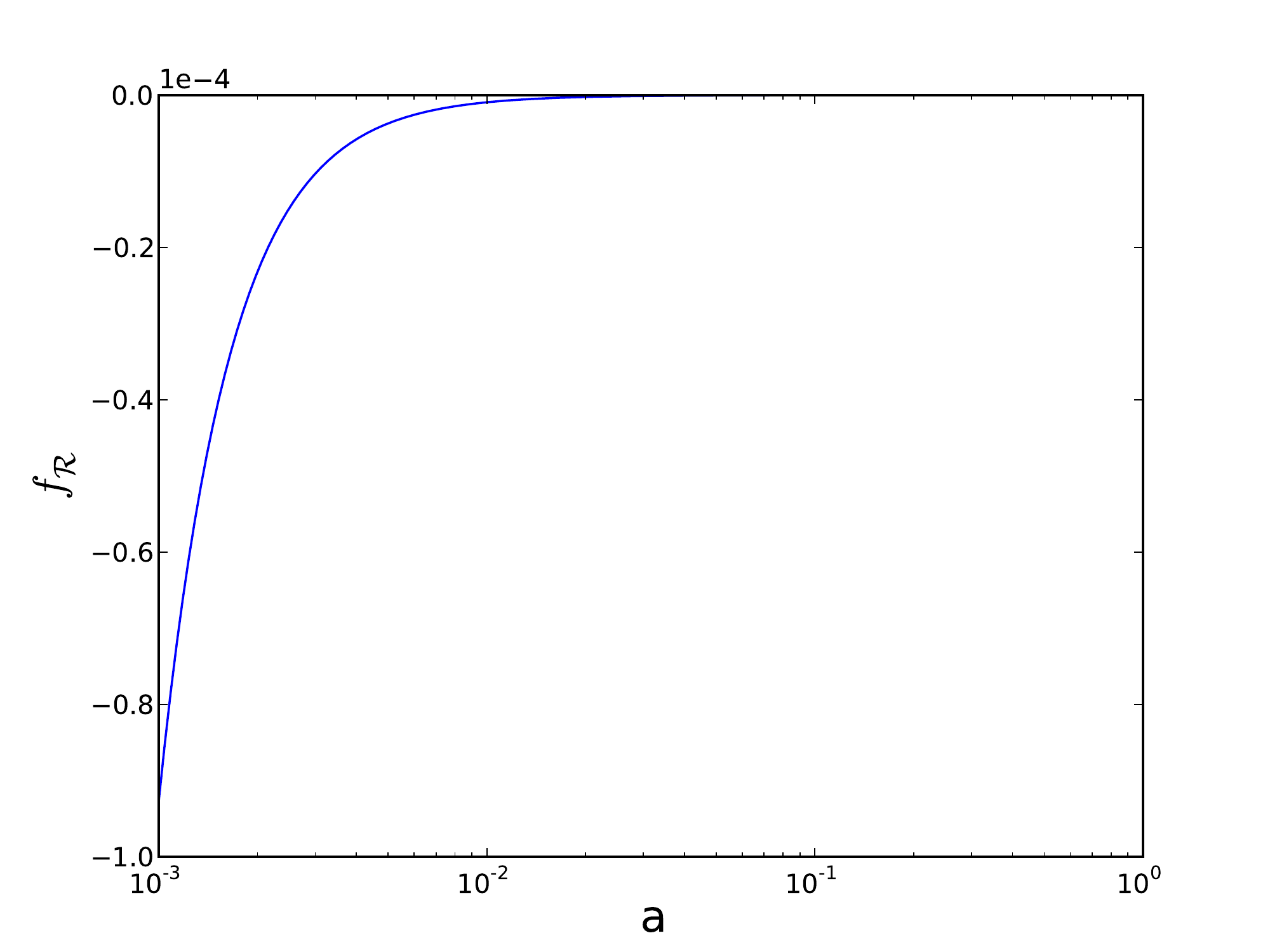} \\
\includegraphics[scale = 0.44]{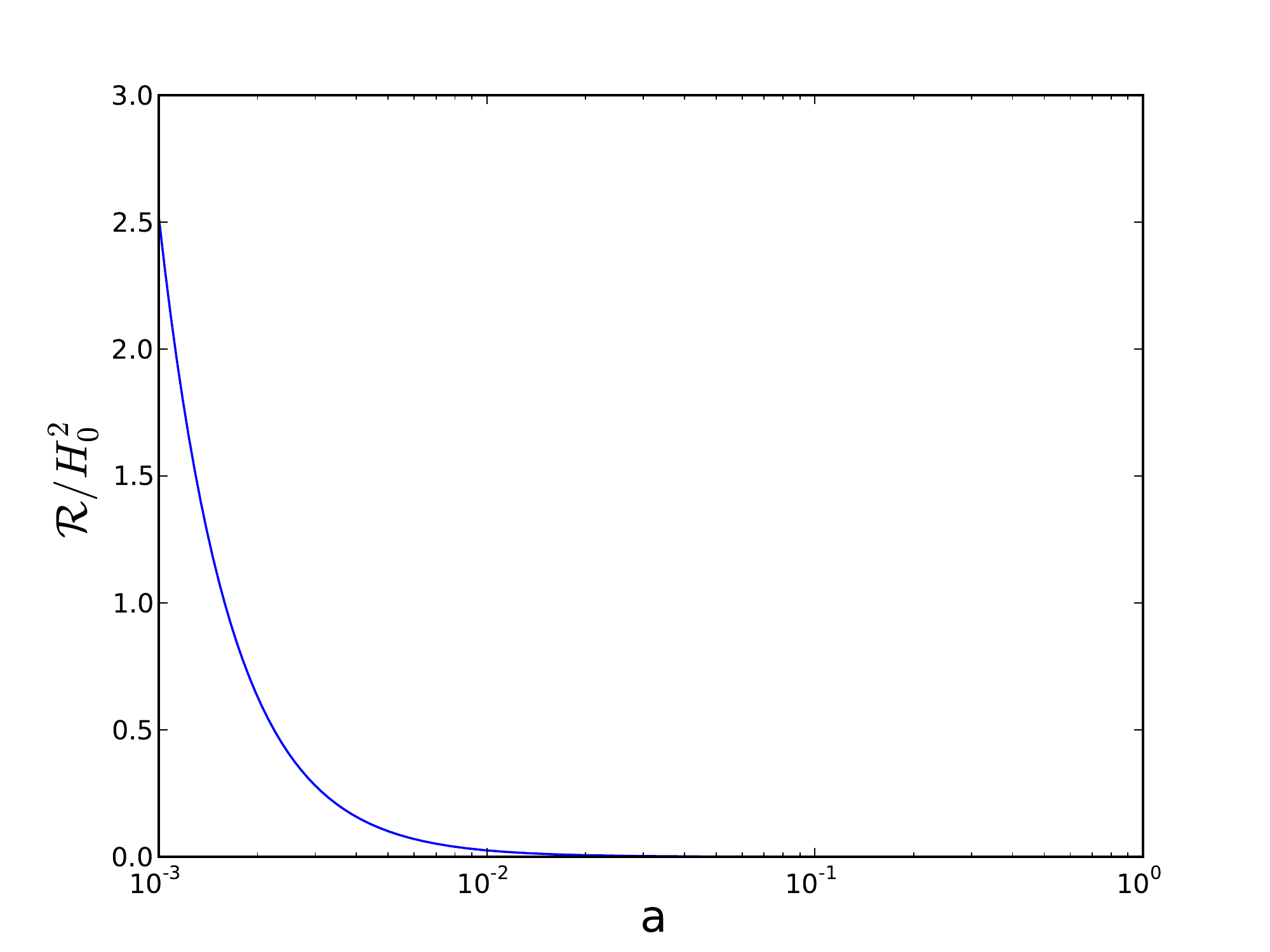}
\end{array}$
\end{center}
\caption{\label{designerevol} Evolution of $f(\mathcal{R})$, $F(\mathcal{R}$) and $\mathcal{R}$, as a function of the scale factor, $a$, obtained using the designer approach for a model with $w_\textrm{eff} = -1$. We have set $C_{1} = -1.0 \times 10^{-8}$ and $C_{2} = -10.5$.}
\end{figure}

In order to recover $f\PC{\mathcal{R}}$, at each step in the evolution, we compute $V\PC{\phi}$ using Eq. (\ref{hubble}). Then, we use the relation between $\mathcal{R}$ and $R$ to compute the first which, with the fact that $V\PC{\phi} = \mathcal{R} F - f\PC{\mathcal{R}}$, lets us determine $f\PC{\mathcal{R}}$. Figure \ref{designerevol} shows the evolution of $f(\mathcal{R})$, $F(\mathcal{R})$ and $\mathcal{R}$ for a model with $C_{1} = -1.0 \times 10^{-8}$ and $C_{2} = -10.5$. This particular choice of values ensures the modifications from GR in the distant past are not that significant, as well ensuring we have a positive definite effective mass for the scalar field.

We can clearly see in Fig. \ref{designerevol} that the modifications from standard GR, even though not large, are more significant in the distant past, as expected, with $f(\mathcal{R})$ hitting a maximum absolute deviation from $\Lambda$ of $10^{-4}$. And, as the evolution progresses, these differences start to vanish, with $f(\mathcal{R})$ tending to the exact Cosmological Constant value, and both $F(\mathcal{R})$ and $\mathcal{R}$ tending to zero.

\section{\label{II}Perturbation theory in the Jordan Frame}

In this section, we derive the equations governing the evolution of scalar perturbations in the hybrid metric-Palatini gravity theory. We will be working in the Jordan frame, following the notation of Kodama and Sasaki \cite{kosa} for general perturbations. We will present the results in conformal time, $\tau$, such that $dt = a \hspace{0.5 mm} d\tau$. Note that, now, $\mathcal{H} = a H$ will be the Hubble parameter in conformal time and an overdot will represent a differentiation with respect to conformal time.

Following Ref. \cite{kosa}, one can separate the spatial and time dependences of the perturbations of the metric. Therefore, for a given wave-number $k$, the metric can be decomposed into four time dependent perturbations $A$, $B$, $H_L$ and $H_T$
\begin{eqnarray}
 g_{00} &=& -a^2\PC{1 + 2AY (k,x)} \nonumber, \\
 g_{0i} &=& -a^2BY_i \nonumber,\\
 g_{ij} &=& a^{2}\PC{\gamma_{ij} + 2 H_L Y \gamma_{ij} + 2 H_T Y_{ij}},
\end{eqnarray}
\noindent where $Y(k,x)$ is the complete set of scalar harmonic functions and $\gamma_{ij}$ is the spatial metric. In this work, we consider the case for which
\begin{eqnarray}
 Y &\propto& e^{i \vec{k} \cdot \vec{x}}, \\
 Y_i &=& - \frac{1}{k}\nabla_{i} Y,
\end{eqnarray}
\noindent and,
\begin{equation}
 Y_{ij} = \PC{k^{-2} \nabla_{i} \nabla_{j} + \frac{\delta_{ij}}{3}}Y.
\end{equation}

On the other hand, the stress-energy tensor perturbations can be decomposed into 4 components: density, $\delta \rho \equiv \rho \delta$; velocity, $v$; isotropic pressure, $\delta p$, and anisotropic stress, $\frac{3}{2}\PC{\rho + p} \sigma$, as
\begin{eqnarray}
 T^{0}_{0} &=& -\rho\PC{1 + \delta Y} \nonumber,\\
 T^{0}_{i} &=& \PC{\rho + p}\PC{v - B}Y_{i} \nonumber,\\
 T^{i}_{j} &=& p \hspace{0.5 mm} \delta^{i}_{j} + \delta p \hspace{0.5 mm} \delta^{i}_{j} + \frac{3}{2}\PC{\rho + p} \sigma Y^{i}_{j},
\end{eqnarray}
\noindent where we have adopted the anisotropic stress notation of Ma and Bertschinger \cite{maber}.

Since there is no dependence of the matter Lagrangian, $\mathcal{L}_{m}$, on the $f(\mathcal{R})$ modifications, in the Jordan frame, the matter conservation equations do not differ from those of standard General Relativity, which are given by
\begin{widetext}
\begin{eqnarray}
\label{pertmatter1}
 &&\dot{\delta} + \PC{1+w}\PC{kv+3\dot{H_{L}}} + 3 \mathcal{H}\PC{\frac{\delta p}{\delta \rho} - w}\delta = 0, \\
 \label{pertmatter2}
 &&\dot{v} - \dot{B} + \mathcal{H}\PC{1-3w}\PC{v - B} + \frac{\dot{w}}{1+w}\PC{v-B} - \frac{\delta p/ \delta \rho}{1+w} k \delta - k A + k \sigma = 0.
\end{eqnarray}
\end{widetext}
\noindent The four perturbed field equations provide two additional independent equations. Here, we present all of the four equations and the perturbed equation of motion of the scalar field for completeness, leaving some intermediate results for Appendix \ref{appendix1}.
\begin{widetext}
\begin{flushleft}
  \underline{0 - 0 component}
\end{flushleft}
\begin{eqnarray}{\label{pert00}}
 &2&\PC{1+\phi}\PR{A\PC{3\mathcal{H}^2 + \frac{3\dot{\phi}^2}{4\phi\PC{1+\phi}} + \frac{3\mathcal{H}\dot{\phi}}{\PC{1+\phi}}} - B\PC{\mathcal{H}k + \frac{k \dot{\phi}}{2\PC{1+\phi}}} - \dot{H_L}\PC{\frac{3\dot{\phi}}{2\PC{1+\phi}} + 3\mathcal{H}} - k^2\PC{H_L + \frac{H_T}{3}}} \nonumber \\
 &-&\delta \phi \PC{k^2 + 3 \mathcal{H}^2 - \frac{3\dot{\phi}^2}{4 \phi^2} - a^2\frac{V_{,\phi}}{2}} - \dot{\delta \phi}\PC{\frac{3\dot{\phi}}{2\phi} + 3\mathcal{H}} = -\kappa^2 a^2 \rho \delta 
\end{eqnarray}
\begin{flushleft}
 \underline{$i - i$ component}
\end{flushleft}
\begin{eqnarray}{\label{pertii}}
 &2&\PC{1+\phi} \Bigg[ A\PC{\mathcal{H}^2 + 2\dot{\mathcal{H}} + \frac{\ddot{\phi} + \dot{\phi}\mathcal{H}}{\PC{1+\phi}} - \frac{3\dot{\phi}^2}{4\phi\PC{1+\phi}} - \frac{k^2}{3}} + \dot{A}\PC{\mathcal{H} + \frac{\dot{\phi}}{2\PC{1+\phi}}} - B\PC{\frac{2k\mathcal{H}}{3} + \frac{k\dot{\phi}}{3\PC{1+\phi}}} - \frac{k\dot{B}}{3} \nonumber \\
 &-& \dot{H_L}\PC{\frac{\dot{\phi}}{\PC{1+\phi}} + 2\mathcal{H}} - \ddot{H_L} - \frac{k^2}{3}\PC{H_L + \frac{H_T}{3}}  \Bigg] - \delta \phi \PC{\frac{3\dot{\phi}^2}{4\phi^2} + \frac{2k^2}{3} + 2\dot{\mathcal{H}} + \mathcal{H}^2 - a^2\frac{V_{,\phi}}{2}} - \dot{\delta \phi}\PC{\mathcal{H} - \frac{3\dot{\phi}}{2\phi}} \nonumber \\
 &-& \ddot{\delta \phi} = \kappa^2 a^2 \delta p
\end{eqnarray}
\begin{flushleft}
  \underline{0 - $i$ component}
\end{flushleft}
 \begin{equation}{\label{pertmomentum}}
  2\PC{1+\phi}\PR{A\PC{\mathcal{H} + \frac{\dot{\phi}}{2\PC{\phi + 1}}} - \dot{H_L} - \frac{\dot{H_T}}{3}} + \delta \phi \PC{\frac{3 \dot{\phi}}{2 \phi} + \mathcal{H}} - \dot{\delta \phi} =  \frac{\kappa^2 a^2}{k}\PC{\rho + p}\PC{v - B}
  \end{equation}
    \begin{flushleft}
  \underline{$i - j$ ($i \neq j$) component}
\end{flushleft} 
 \begin{equation}{\label{pertaniso}}
  \PC{1+\phi}\PR{-k^2 A - k\PC{\dot{B} + 2\mathcal{H}B} + \ddot{H_T} + 2\mathcal{H}\dot{H_T} - k^2\PC{H_L + \frac{H_T}{3}}} - k^2 \delta \phi - \dot{\phi}\PC{kB-\dot{H_T}} = \frac{3}{2}\kappa^2 a^2 \PC{\rho + p} \sigma
  \end{equation}
  \begin{flushleft}
   \underline{$\delta \phi$ equation}
  \end{flushleft}
 \begin{eqnarray}{\label{pertscalar}}
  &&\ddot{\delta \phi} + \dot{\delta \phi}\PC{2 \mathcal{H} - \frac{\dot{\phi}}{\phi}} + \delta \phi \PC{k^2 + \frac{\dot{\phi}^2}{2 \phi^2} - \frac{2}{3} a^2 \phi V_{,\phi \phi} + a^2\frac{R}{3}} + A\PC{\frac{\dot{\phi}^2}{\phi} - 2\ddot{\phi} - 4\mathcal{H}\dot{\phi}} \nonumber \\
  && + \dot{\phi}\PC{3 \dot{H_L} - \dot{A} + k B } = -a^2\frac{\phi}{3}\delta R,
 \end{eqnarray}
\noindent where $\delta R$ is the perturbed Ricci scalar, given by
  \begin{equation}{\label{pertricci}}
   \delta R = \frac{2}{a^2}\PR{-6 \frac{\ddot{a}}{a}A - 3\mathcal{H}\dot{A} + k^2 A + k\dot{B} + 3 k \mathcal{H}B + 9 \mathcal{H} \dot{H_L} + 3 \ddot{H_L} + 2k^2 \PC{H_L + \frac{H_T}{3}}}Y.
  \end{equation}

\end{widetext}

\subsection{\label{Newtonian}Conformal Newtonian Gauge}

In this particular gauge, one sets $H_T = B = 0$, $A = \Psi$ and $H_L = -\Phi$, following the notation of Ma and Bertschinger \cite{maber}. In this gauge, one can combine equations (\ref{pertmatter1}) and (\ref{pertmatter2}) for a Cold Dark Matter (CDM) and radiation overdensity, providing a set of single second--order differential equations, given by
\begin{eqnarray}
 &&\ddot{\delta_{c}} + \mathcal{H}\dot{\delta_c} + k^2 \Psi - 3 \ddot{\Phi} - 3 \mathcal{H} \dot{\Phi} = 0,\\
 &&\ddot{\delta_{\gamma}} + \frac{1}{3}k^2 \delta_{\gamma} + \frac{4}{3}k^2 \Psi - 4\ddot{\Phi} = 0, 
\end{eqnarray}
\noindent respectively. The Einstein equations, on the other hand, in this gauge, which have already been presented in Ref. \cite{hybridcosmo}, are given by
\begin{widetext}
\begin{flushleft}
 \underline{$0-0$ component}
\end{flushleft}
\begin{eqnarray}{\label{00newtonian}}
 &&2\PC{1+\phi}\PR{ \Psi \PC{3 \mathcal{H}^2 + \frac{3 \dot{\phi}^2}{4\phi\PC{1+\phi}} + \frac{3 \mathcal{H}\dot{\phi}}{\PC{1+\phi}}} + \dot{\Phi}\PC{3 \mathcal{H} + \frac{3 \dot{\phi}}{2\PC{1+\phi}}} + k^2 \Phi   } - \dot{\delta \phi}\PC{\frac{3\dot{\phi}}{2\phi} + 3 \mathcal{H}} \nonumber \\
&&-\delta \phi \PC{k^2 + 3\mathcal{H}^2 - \frac{3 \dot{\phi}^2}{4\phi^2} - a^2 \frac{V_{,\phi}}{2}} = -\kappa^2 a^2 \rho \delta 
\end{eqnarray}
\begin{flushleft}
 \underline{$i-i$ component}
\end{flushleft}
\begin{eqnarray}{\label{iinewtonian}}
&&2\PC{1+\phi} \PR{ \Psi \PC{ \mathcal{H}^2 + 2 \dot{\mathcal{H}} + \frac{\ddot{\phi} + \mathcal{H}\dot{\phi}}{\PC{1+\phi}} - \frac{3\dot{\phi}^2}{4\phi \PC{1+\phi}}} + \frac{k^2}{3} \PC{\Phi-\Psi} + \dot{\Psi}\PC{\mathcal{H} + \frac{\dot{\phi}}{2\PC{1+\phi}}} + \dot{\Phi} \PC{\frac{\dot{\phi}}{\PC{1+\phi}} + 2\mathcal{H}} + \ddot{\Phi} } \nonumber \\
&& - \delta \phi \PC{\frac{3 \dot{\phi}^2}{4 \phi^2} + \frac{2}{3}k^2 + 2 \dot{\mathcal{H}} + \mathcal{H}^2 - a^2 \frac{V_{,\phi}}{2}} - \dot{\delta \phi} \PC{\mathcal{H} - \frac{3 \dot{\phi}}{2\phi}} - \ddot{\delta \phi} = \kappa^2 a^2 \delta p
\end{eqnarray}
\begin{flushleft}
 \underline{$0-i$ component}
\end{flushleft}
 \begin{eqnarray}{\label{momentumnewton}}
 2\PC{1 + \phi} \PR{\Psi \PC{\mathcal{H} + \frac{\dot{\phi}}{2 \PC{\phi+1}}} + \dot{\Phi}} + \delta \phi \PC{\mathcal{H} + \frac{3 \dot{\phi}}{2 \phi}} - \dot{\delta \phi} = \frac{\kappa ^2 a^2}{k} \PC{\rho + p} v
 \end{eqnarray}
 \begin{flushleft}
  \underline{$i$ - $j$ ($i \neq j$) component}
 \end{flushleft}
\begin{eqnarray}{\label{ijnewtonian}}
 \PC{1+\phi} \PC{k^2 \Phi -k^2 \Psi} -k^2 \delta \phi = \frac{3}{2}\kappa^2 a^2 \Sigma \PC{\rho + p} \sigma.
\end{eqnarray}
\end{widetext}
\noindent The last equation we show is the perturbed Klein-Gordon equation for the scalar field,
\begin{widetext}
\begin{flushleft}
 \underline{$\delta \phi$ equation}
\end{flushleft}
\begin{eqnarray}{\label{scalarnewton}}
 &&\ddot{\delta \phi} + \dot{\delta \phi} \PC{2 \mathcal{H} - \frac{\dot{\phi}}{\phi}} + \delta \phi \PC{k^2 + \frac{\dot{\phi}^2}{2 \phi^2} - \frac{2}{3} a^2 \phi V_{,\phi \phi} + a^2\frac{R}{3}} + \Psi \PC{ \frac{\dot{\phi}^2}{\phi} - 2 \ddot{\phi} - 4 \dot{\phi} \mathcal{H} } \nonumber \\
 &&- \dot{\phi}\PC{3\dot{\Phi} + \dot{\Psi}} = -\frac{\phi}{3}a^2 \delta R,
\end{eqnarray}
\begin{flushleft}
 \noindent where
\end{flushleft}
\begin{equation}{\label{deltarnewton}}
 \delta R = \frac{2}{a^2}\PR{-6 \frac{\ddot{a}}{a}\Psi - 3\mathcal{H}\dot{\Psi} + k^2 \Psi - 9 \mathcal{H} \dot{\Phi} - 3 \ddot{\Phi} - 2k^2 \Phi}Y.
\end{equation}

\end{widetext}

\subsection{\label{synchronous}Synchronous Gauge}

In this gauge, $A = B = 0$ and, adopting the notation of Ma and Bertschinger \cite{maber}, $H_L = h/6$ and $H_T = -3\PC{\eta + h/6}$. One may remove the remaining freedom and completely define the synchronous coordinates by setting that cold dark matter particles are at rest in this gauge, having zero peculiar velocity, $v_m$. Hence, the perturbed CDM and radiation evolution equations are written as
\begin{eqnarray}
 &&\dot{\delta_{c}} = -\frac{1}{2}\dot{h},\\
 &&\ddot{\delta_{\gamma}} + \frac{k^2}{3} \delta_{\gamma} - \frac{4}{3} \ddot{\delta_c} = 0,
\end{eqnarray}
\noindent respectively, while the perturbed Einstein equations and the perturbed field equation are given by
\begin{widetext}
\begin{flushleft}
\underline{$0-0$ component}
\end{flushleft}
 \begin{eqnarray}{\label{00sync}}
 2\PC{1+\phi}\PR{k^2 \eta - \frac{\dot{h}}{6}\PC{3\mathcal{H} + \frac{3\dot{\phi}}{2\PC{1+\phi}}}} - \delta \phi \PC{k^2 + 3 \mathcal{H}^2 - \frac{3\dot{\phi}^2}{4\phi^2} - a^2 \frac{V_{,\phi}}{2}} - \dot{\delta \phi} \PC{3\mathcal{H} + \frac{3\dot{\phi}}{2\phi}} = -\kappa^2 a^2 \rho \delta
 \end{eqnarray}
\begin{flushleft}
 \underline{$i-i$ component}
\end{flushleft}
\begin{eqnarray}{\label{iisync}}
 &&2\PC{1+\phi}\PR{-\frac{\dot{h}}{6}\PC{2\mathcal{H} + \frac{\dot{\phi}}{\PC{1+\phi}}} - \frac{\ddot{h}}{6} + \frac{1}{3}k^2 \eta} - \delta \phi \PC{ \frac{3\dot{\phi}^2}{4 \phi^2} + \frac{2 k^2}{3} + 2 \dot{\mathcal{H}} + \mathcal{H}^2 - a^2 \frac{V_{,\phi}}{2} } - \dot{\delta \phi}\PC{\mathcal{H} - \frac{3 \dot{\phi}}{2\phi}} - \ddot{\delta \phi} \nonumber \\
 &&= \kappa^2 a^2 \delta p
\end{eqnarray}
\begin{flushleft}
 \underline{$0-i$ component}
\end{flushleft}
\begin{eqnarray}{\label{momentumsync}}
 2\PC{1+\phi}\dot{\eta} + \delta \phi \PC{\mathcal{H} + \frac{3\dot{\phi}}{2\phi}} - \dot{\delta \phi} = \frac{\kappa^2 a^2}{k} \PC{\rho + p} v
\end{eqnarray}
\begin{flushleft}
 \underline{$i-j$ ($i\neq j$) component}
\end{flushleft}
\begin{eqnarray}{\label{ijsync}}
 \PC{1+\phi}\PR{k^2 \eta - 6\mathcal{H}\PC{\dot{\eta} + \frac{\dot{h}}{6}} -3 \ddot{\eta} - \frac{\ddot{h}}{2}} - 3\dot{\phi}\PC{\dot{\eta} + \frac{\dot{h}}{6}} - k^2 \delta \phi = \frac{3}{2} \kappa^2 a^2 \PC{\rho + p} \sigma. 
\end{eqnarray}
\begin{flushleft}
 \underline{$\delta \phi$ equation}
\end{flushleft}
\begin{eqnarray}{\label{deltaphisync}}
 \ddot{\delta \phi} + \dot{\delta \phi}\PC{2 \mathcal{H} - \frac{\dot{\phi}}{\phi}} + \delta \phi \PC{k^2 + \frac{\dot{\phi}^2}{2 \phi^2} - \frac{2}{3} a^2 \phi V_{,\phi \phi} + a^2\frac{R}{3}} + \dot{\phi} \frac{\dot{h}}{2} = -a^2 \frac{\phi}{3}\delta R,
\end{eqnarray}
\end{widetext}
 \noindent where $\delta R$ is given by
\begin{equation}{\label{deltarsync}}
 \delta R = \frac{2}{a^2}\PR{\frac{3}{2} \mathcal{H}\dot{h} + \frac{\ddot{h}}{2} - 2 \eta k^2}Y.
\end{equation}

\section{\label{III}The Lensing Potential}

In this section, we will be working in the Newtonian gauge with the anisotropy and Poisson equations. The former establishes a relation between the Newtonian potentials $\Psi$ and $\Phi$, while the latter describes the dependence of $\Phi$, the curvature potential, on the matter comoving density perturbation
\begin{equation}{\label{cosmoving}}
 \Delta = \delta + 3\frac{aH}{k} \mathcal{V},
\end{equation}
\noindent where $\mathcal{V} \equiv \PC{1+w}v$. 

As shown in Section \ref{Newtonian}, the anisotropy equation for theories described by the action (\ref{scalaraction}) is given by Eq. (\ref{ijnewtonian}). In order to obtain the Poisson equation, one has to write Eq. (\ref{pertmomentum}) in the Newtonian gauge and combine it with Eq. (\ref{00newtonian}), the result being
\begin{widetext}
\begin{eqnarray}{\label{Poisson}}
 \frac{k^2}{a^2 H^2} \Phi = &-&\frac{3}{2}\frac{H_{0}^{2}}{D \PC{\phi} H^2} E_i \Delta_i - 3\Psi \PC{\frac{\phi^{\prime 2}}{4\phi D \PC{\phi}} + \frac{\phi^{\prime}}{2 D \PC{\phi}}} - \frac{3}{2}\frac{\Phi^{\prime} \phi^{\prime}}{D \PC{\phi}} + \frac{\delta \phi}{2 D\PC{\phi}} \PR{\frac{k^2}{a^2 H^2} + 3 \PC{2 - \frac{\phi ^{\prime2}}{4 \phi^2} + \frac{3 \phi^{\prime}}{2\phi} } - \frac{V_{, \phi}}{2 H^2}} \nonumber \\
 &+& \frac{3}{4} \frac{\phi^{\prime} \delta \phi^{\prime}}{\phi D \PC{\phi}},
\end{eqnarray}
\end{widetext}
\noindent where we have defined, for simplification, $D\PC{\phi} \equiv 1 + \phi$, and introduced $E_i = \kappa ^2\rho_i/3 H_{0}^{2}$. The repeated indices represent a sum over the matter fields, and the prime indicates a differentiation with respect to $\ln a$.

Neglecting anisotropic contributions from matter fields (i.e.~setting $\sigma_i = 0$), eq. (\ref{ijnewtonian}) yields the following relation between the gravitational potentials
\begin{equation}{\label{anisotropy}}
 \Phi - \Psi = \frac{\delta \phi}{D\PC{\phi}}.
\end{equation}
\noindent This stands as a clear departure from standard General Relativity where the anisotropy equation would be a simple constraint, $\Psi = \Phi$, reducing the number of independent perturbed variables. On the other hand, the Poisson equation would also just be an algebraic relation between the comoving matter density perturbation $\Delta_i$ and the curvature perturbation $\Phi$. 

However, in the hybrid metric-Palatini gravity, Eqs. (\ref{Poisson}) and (\ref{anisotropy}) have extra dynamics. In the anisotropy equation, these are encoded in the slip between the Newtonian potentials. Following the work done in Refs. \cite{frperturbed,frmine} for $f(R)$ theories, we will choose to evolve the slip as one of the perturbed variables. And, alongside it, we choose to evolve another function of the Newtonian potentials $\Psi$ and $\Phi$, which facilitates the numerical treatment of the equations and the interpretation of the results. Namely, we evolve the following variables
\begin{eqnarray}
\label{lensing}
 &&\Phi_{+} = \frac{\Phi + \Psi}{2} \\
 &&\chi \equiv \delta \phi = D\PC{\phi} \PC{\Phi - \Psi},
\end{eqnarray}
\noindent along with the perturbations of matter fields. Now, eq. (\ref{anisotropy}) becomes a constraint equation and any non-zero value of $\chi$ will signal a deviation from standard GR. The variable $\Phi_{+}$ is the lensing potential, i.e. the combination of potentials responsible for such effects as the integrated Sachs--Wolfe effect in the CMB and weak lensing of distant galaxies. Ignoring any contribution from the radiation component, the momentum (\ref{momentumnewton}) and the Poisson (\ref{Poisson}) equations provide a set of coupled first--order differential equations for $\Phi_{+}$ and $\chi$, given by
\begin{widetext}
 \begin{eqnarray}
  \label{lensingevolution}
  &&\Phi_{+}^{\prime} = \frac{3}{2} \frac{a H_{0}^{2}}{H k D} E_i \mathcal{V}_i - \Phi_{+}\PC{1 + \frac{D^{\prime}}{2D}} - \frac{3}{4} \frac{D^\prime \chi}{\phi D^2} \\
  \label{slipevolution}
  &&\chi^{\prime} = 2 \frac{H_{0}^{2}}{H^2} \frac{\phi D}{D^{\prime}} E_i \Delta_i + 2\Phi_{+}^{\prime}\phi D - 4 \chi \PR{  \frac{\phi D}{D^{\prime}} + \frac{D^{\prime}}{8 \phi D} \PC{  2\phi^2 - D} + \frac{1}{4}\PC{3D + \phi} - \frac{V_{,\phi}}{12 H^2} \frac{\phi D}{D^{\prime}} } \nonumber \\
&& \hspace{0.75 cm} +  2\Phi_{+}\PC{\frac{D D^{\prime}}{2} + \phi D + \frac{2}{3}\frac{k^2}{a^2 H^2} \frac{\phi D^2}{D^{\prime}}} 
 \end{eqnarray}
\end{widetext}

\begin{figure}[t!]
\begin{center}$
\begin{array}{c}
\includegraphics[scale = 0.425]{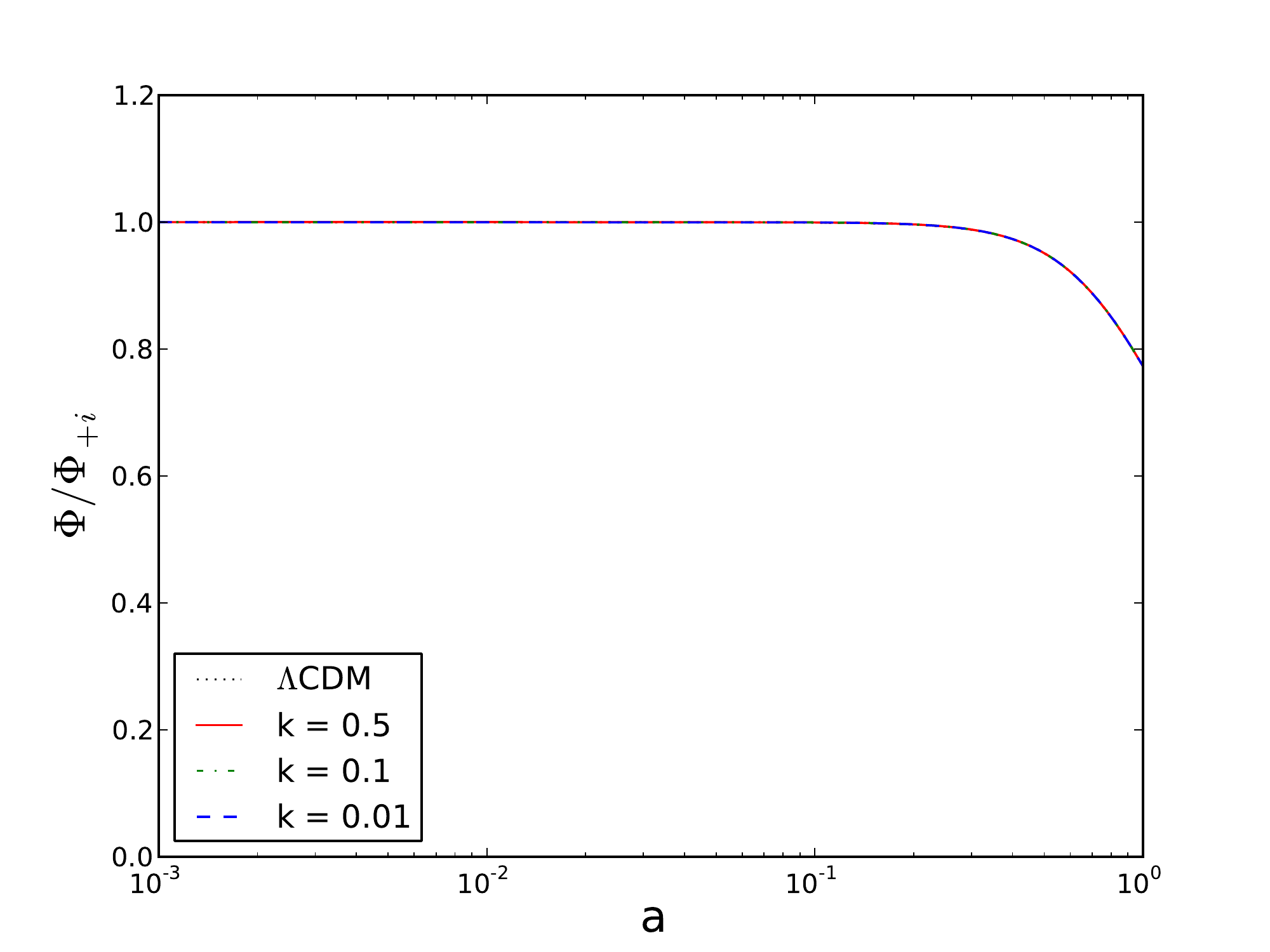} \\
\includegraphics[scale = 0.425]{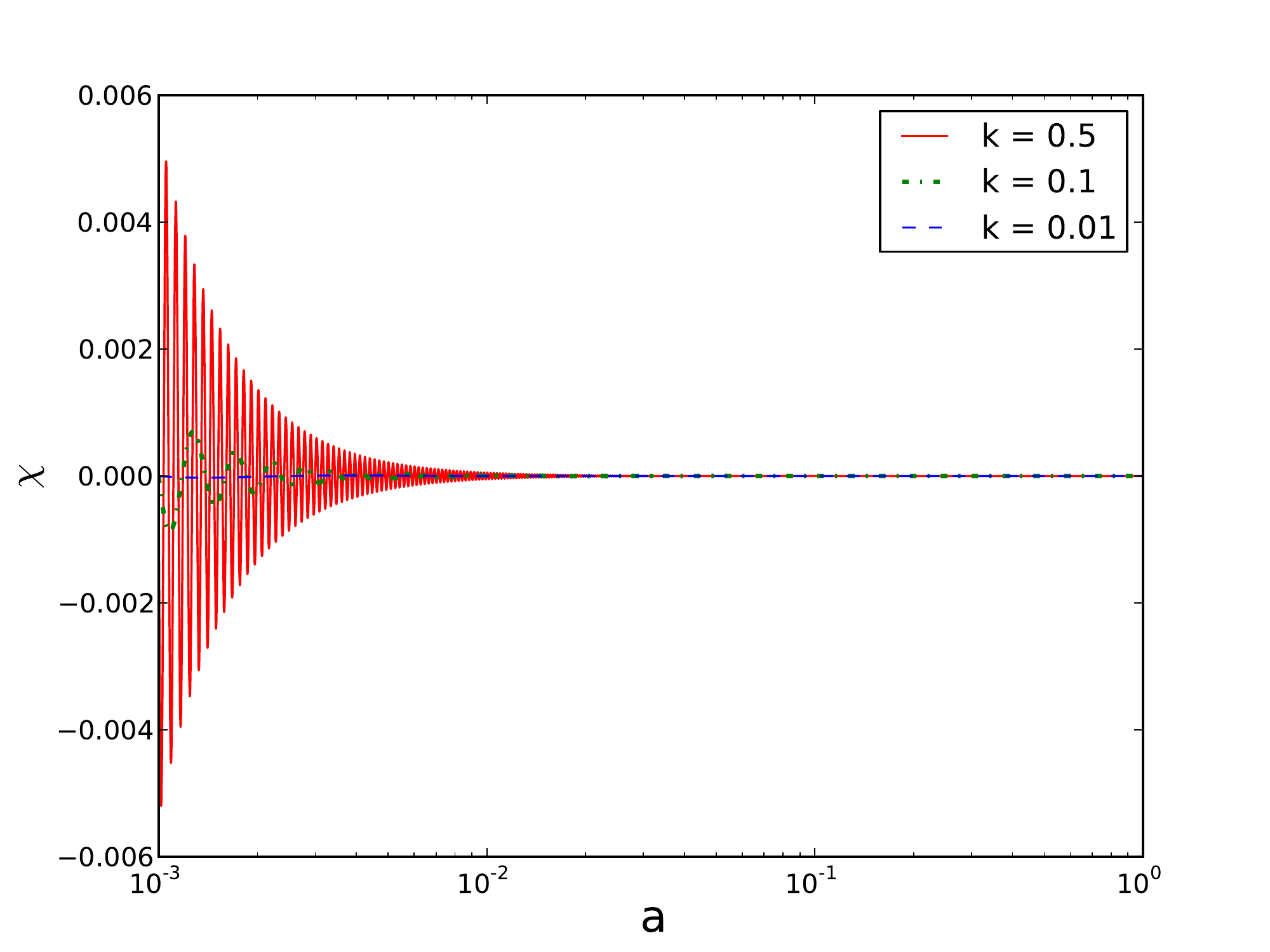} \\
\includegraphics[scale = 0.425]{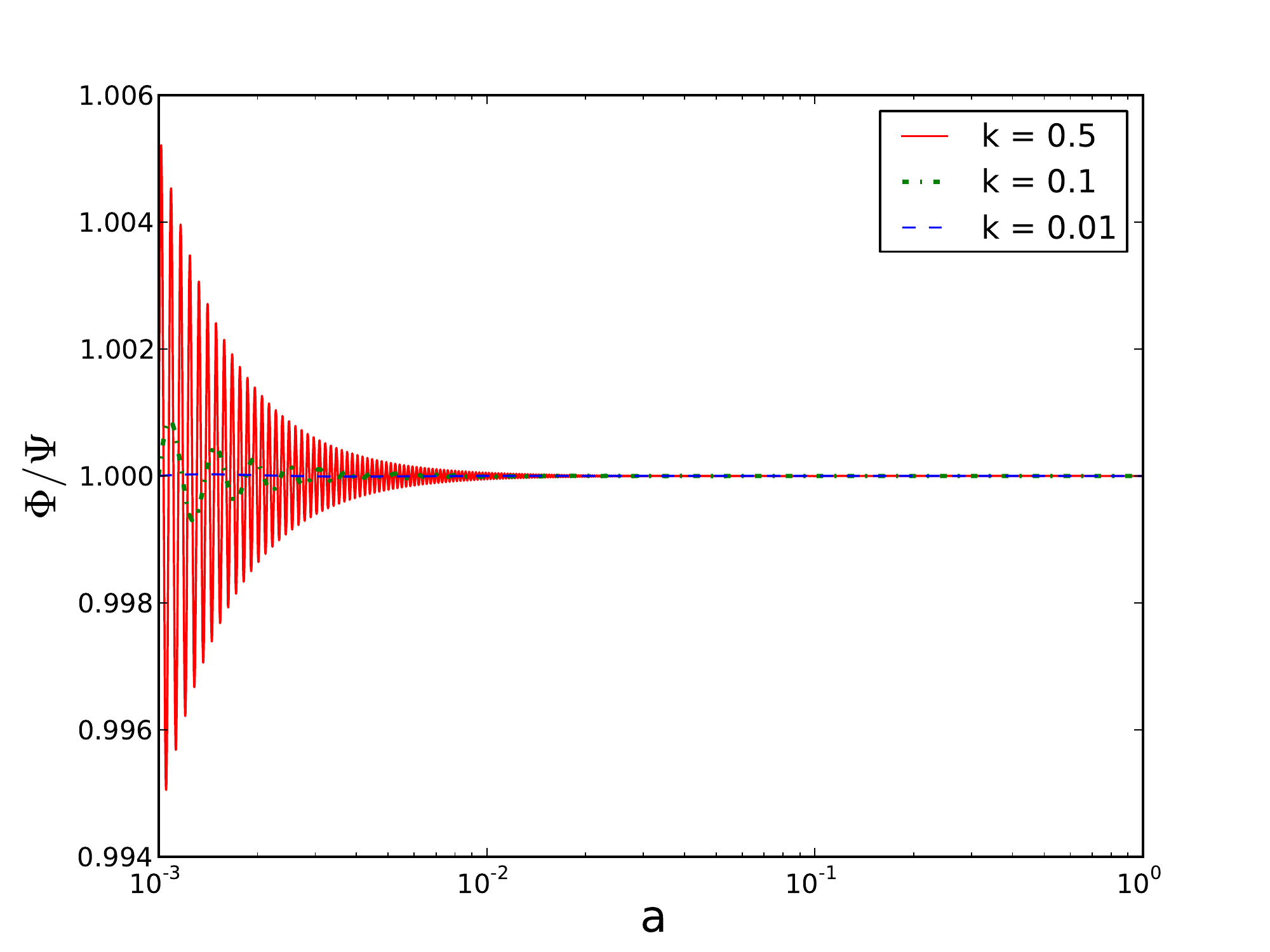}
\end{array}$
\end{center}
\caption{\label{figure_pots} The evolution of the lensing potential, $\Phi_{+}$, $\chi$ and the ratio between the Newtonian potentials, $\Phi$ and $\Psi$, as a function of the scale factor, $a$, for the designer model with $w_{\textrm{eff}} = -1$ under analysis. Note that, for $\Lambda$CDM, $\chi = 0$ and $\Phi/\Psi = 1$ throughout the entire evolution. We plot this for three different $k$ ($h$/Mpc) modes.}
\end{figure}

\begin{figure}[t!]
 \begin{center}$
 \begin{array}{c}
  \includegraphics[scale = 0.45]{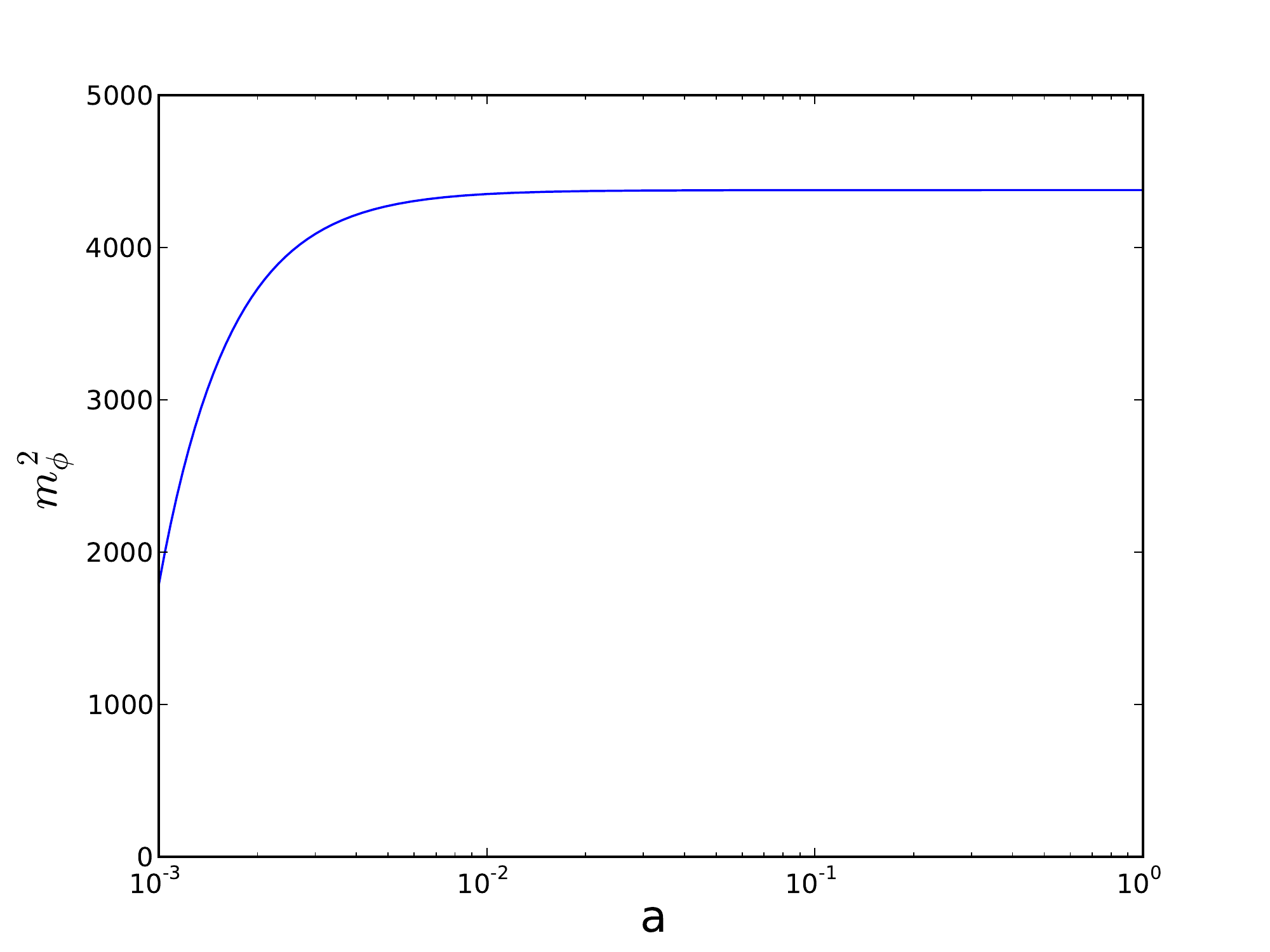}
  \end{array}$
 \end{center}
 \caption{\label{scalarmass} The evolution of the effective scalar mass, $m^{2}_{\phi}$, for the designer model $w_{\textrm{eff}} = -1$ considered in this analysis.}
\end{figure}

The numerical evolution of the above equations is shown in Figure \ref{figure_pots}, for a model with $w_{\textrm{eff}} = -1$ and a value of $|F(\mathcal{R})| \approx 10^{-4}$ at the start of evolution, which we have set at $z_i = 1000$. This ensures that, at this point, the deviations from General Relativity are small and we can set the initial conditions as if that was the case. Therefore, we set $\Phi_{+}(z = z_i) = -1$ and $\chi (z = z_i) = 0$ and use the standard GR relations for the matter perturbations $v_m$ and $\Delta_m$,
\begin{eqnarray}
 && v_{m} (z = z_i) = \frac{2 k}{3 a H}\Phi_{+}, \\
 && \Delta_{m} (z = z_i) = - \frac{2 k^2}{3 a^2 H^2}\Phi_{+}.
\end{eqnarray}
In Fig. \ref{figure_pots}, one can clearly see oscillations in $\chi$. This is expected since equation (\ref{scalarnewton}) resembles that of a damped oscillator, whose frequency of oscillation will increase with $k$. Also, we note that the deviations from $\Lambda$CDM are actually larger in the distant past. This is due to the behavior of the effective mass of the scalar field, $m_{\phi}$, whose evolution for the model under analysis is plotted in Fig. \ref{scalarmass}. The latter, in turn, defines the range of action of the modifying fifth force: if we were to define a Compton wavelength for its range, this would be inversely proportional to $m_{\phi}$. Hence, for $a \ll 1$, the range of action of the fifth force is quite significant, leading to a greater amplitude in the oscillations of $\chi$, which is then damped with the evolution.

Of course, the amplitude of the slip oscillations is larger for the smaller scales (or larger $k$) as these begin their evolution deep within the range of action of the fifth force. Nevertheless, neither the enhancement or the oscillations in the slip are reflected on $\Phi_{+}$, which remains almost indistinguishable from $\Lambda$CDM throughout the entire evolution, apart from a negligible increment at the beginning of it, for every scale considered. These oscillations in $\chi$ do, however, translate into rapid oscillations in the ratio between the Newtonian potentials, $\Phi$ and $\Psi$, which could potentially be discriminating between this particular model and GR.

\section{\label{conclusion}Conclusion}

In this paper, we have derived the full set of perturbed Einstein equations for the novel hybrid metric-Palatini theory of gravity and presented them in the Newtonian and synchronous gauges. The latter, in particular, open the possibility of implementing this model in CAMB \cite{camb} and give an in-depth analysis of the effects it can have at early times and even constrain its parameter space, which we leave for future work.

We have introduced a designer approach to obtain a family of functions $f(\mathcal{R})$ that reproduce a cosmology indistinguishable from $\Lambda$CDM, with an effective equation of state exactly equal to $w_{\textrm{eff}} = -1$. This particular approach leads to models where the modifications from standard General Relativity are more significant in the distant past. And, even though one can tweak the free parameters to control such modifications in order that they are quite negligible at a redshift of $z_{i} \approx 1000$, this can prove problematic at even earlier times if the departure from GR gets increasingly larger, as it seems to do. Potentially, one could observe an inversion in the sign of $G_{\textrm{eff}}$, leading to an inversion of the effect of gravity. This was, however, avoided in our analysis. We would like to point out that, other background solutions were neglected, for now, due to the inability to consistently set the initial conditions for $F$ when $w_{\textrm{eff}} \neq -1$. This is 
being explored for future work.

We also derived the Poisson equation, which is substantially different from standard GR, given the inclusion of several extra dynamical elements. We then introduced the lensing potential, $\Phi_{+}$, and the slip between the Newtonian potentials, $\chi$, which we numerically evolved using the designer approach. We note that the departure from GR is more noticeable at the beginning of evolution. More specifically, $\chi$ oscillates with a frequency proportional to the mode's wave number, and the oscillations' amplitude is the largest at this point. It is then gradually damped due to the evolution, tending to a GR value of $0$.

Nevertheless, these oscillations never end up reflecting upon the lensing potential, $\Phi_{+}$, which remains practically indistinguishable from $\Lambda$CDM apart from a negligible enhancement at the start of evolution. However, they do reflect upon the Newtonian potentials, translating into a signature on the ratio between them, which oscillates rather quickly, signaling a clear departure of this model from standard GR.

We also note that the evolution of $\chi$ and $\Phi_{+}$ in the hybrid metric-Palatini theory is related, like in metric $f(R)$ models, to the effective mass of the additional scalar degree of freedom, $m^{2}_{\phi}$. Since the latter is smaller at early times, the range of the action of the additional fifth force will be larger. Hence, the enhancement in the perturbations, specially $\chi$, will be greater then. And it will be greater the smaller the scale under consideration is, since these scales start their evolution deep within the range of the additional force.

\acknowledgments

The author would like to thank Andrew Liddle for invaluable disussions and comments on this paper. N.A.L.\ also acknowledges financial support from Funda\c{c}\~{a}o para a Ci\^{e}ncia e a Tecnologia (FCT) through grant SFRH/BD/85164/2012.

\bibliography{hybrid}

\appendix

\section{Jordan frame perturbation equations}{\label{appendix1}}

For completeness, we show here some of the components used for deriving the results presented in Section \ref{II}. The perturbations to the geometric quantities are unmodified in the hybrid metric-Palatini theory. Hence, as in standard GR, for the Christoffel symbols we have:
\begin{widetext}
\begin{eqnarray}{\label{christoffel}}
 \delta \Gamma_{00}^{0} &=& \dot{A} Y, \hspace{0.25 cm}  \delta \Gamma_{0j}^{0} = -\PR{kA +\mathcal{H}B}Y_{j}, \\
 \delta \Gamma_{ij}^{0} &=& \PR{-2 \mathcal{H}A + \frac{k}{3}B + 2\mathcal{H}H_L + \dot{H_L}} \gamma_{ij} Y + \PR{-kB + 2\mathcal{H} H_T + \dot{H_T}} Y_{ij}, \\
 \delta \Gamma_{00}^{j} &=& -\PR{kA + \dot{B} + \mathcal{H}B} Y^{j}, \\
 \delta \Gamma_{0j}^{i} &=& \dot{H_L} \delta^{i}_{j} Y + \dot{H_T} Y^{i}_{j}, \\
 \delta \Gamma_{jk}^{i} &=& -k H_L \PC{\delta^{i}_{j} Y_k + \delta^{i}_{k} Y_j - \delta_{jk}Y^{i}} + \mathcal{H}B \gamma_{jk} Y^{i} + H_T \PC{Y^{i}_{j , k} + Y^{i}_{k , j} - Y_{jk}^{, i}}. 
\end{eqnarray}
\end{widetext}
\noindent As for the Ricci scalar and the Ricci tensor, we have
\begin{widetext}
\begin{eqnarray}
\label{riccisjordan}
&&\delta R = \frac{2}{a^2}\PR{-6 \frac{\ddot{a}}{a} A - 3\mathcal{H}\dot{A} + k^2A + k\dot{B} + 3k\mathcal{H}B + 9\mathcal{H}\dot{H_L} + 3\ddot{H_L} + 2k^2\PC{H_L + \frac{H_T}{3}}} Y, \\
\label{riccitensor}
&&\delta R_{00} = - \PR{k^2A - 3\mathcal{H}\dot{A} + k\PC{\dot{B} + \mathcal{H}B} + 3\ddot{H_L} + 3\mathcal{H}\dot{H_L}}Y, \\
&&\delta R_{jk} = \Bigg[ -2A\PC{\frac{\ddot{a}}{a} + \mathcal{H}^2} - \mathcal{H}\dot{A} + \frac{k^2}{3}A + \frac{k}{3}\PC{\dot{B} + 5\mathcal{H}B} + \ddot{H_L} + 5\mathcal{H}\dot{H_L} + 2\PC{\frac{\ddot{a}}{a} + \mathcal{H}^2}H_L + \frac{4k^2}{3}\PC{H_L + \frac{H_T}{3}} \Bigg]\delta_{jk}Y \nonumber \\
&&+\PR{-k^2A - k\PC{\dot{B} + \mathcal{H}B} + \ddot{H_T} + \mathcal{H}\dot{H_T} + 2\PC{\frac{\ddot{a}}{a}+\mathcal{H}^2}H_T - k^2\PC{H_L + \frac{H_T}{3}} + \mathcal{H}\PC{\dot{H_T} - kB}}Y_{jk}, \\
&&\delta R_{0j} = \PR{-\PC{\frac{\ddot{a}}{a} + \mathcal{H}^2}B - 2k\mathcal{H}A + 2k\dot{H_L} + \frac{2}{3}k\dot{H_T}}Y_j
\end{eqnarray}
\end{widetext}
\noindent The main additional components come from the perturbations to the covariant derivates of $\phi$ that appear in eq. (\ref{einsteinscalar2})
\begin{widetext}
 \begin{eqnarray}{\label{covarpert}}
&&\delta \PC{\nabla_{\mu} \nabla_{\nu} \phi} = \nabla_{\mu}\nabla_{\nu} \delta \phi - \delta \Gamma_{\mu \nu}^{\alpha} \partial_{\alpha} \phi \\
&&\delta \PC{\nabla^{\mu} \nabla_{\nu} \phi} = \nabla^{\mu}\nabla_{\nu} \delta \phi - \delta g^{\mu \alpha} \nabla_{\alpha} \nabla_{\nu} \phi - g^{\mu \alpha} \delta \Gamma^{\beta}_{\alpha \nu} \partial_{\beta} \phi
 \end{eqnarray} 
\end{widetext}
\noindent We are considering our scalar degree of freedom to be a function such that $\phi = \phi \PC{t} + \delta \phi \PC{x,t}$, and so we get
\begin{widetext}
\begin{eqnarray}
 &&\delta \PC{\nabla^{i} \nabla_{i} \phi} = \frac{Y}{a^2}\PR{-k^2 \delta \phi - 3\mathcal{H} \dot{\delta \phi} + \dot{\phi}\PC{6 \mathcal{H}A - kB - 3\dot{H_L}}} \\
 &&\delta \PC{\nabla^{0} \nabla_{0} \phi} = \frac{Y}{a^2}\PR{-\ddot{\delta \phi} + \mathcal{H} \dot{\delta \phi} + 2 \ddot{\phi}A - 2A\mathcal{H}\dot{\phi} + \dot{\phi} \dot{A}} \\
 &&\delta \PC{\nabla^{0} \nabla_{i} \phi} = \frac{Y_i}{a^2}\PR{k \dot{\delta \phi} - \mathcal{H} k \delta \phi - kA \dot{\phi}} \\
 &&\delta \PC{\nabla^{i} \nabla_{k} \phi} = \frac{Y^{i}_{k}}{a^2}\PR{k^2 \delta \phi + \dot{\phi} \PC{kB - \dot{H_T}}} \\
 &&\delta \PC{\nabla_{i} \nabla_{j} \phi} = \frac{\delta_{ij} Y}{a^2}\PR{-\frac{k^2}{3}\delta \phi - \mathcal{H} \dot{\delta \phi} + \dot{\phi}\PC{2\mathcal{H}A-\frac{k}{3}B-2\mathcal{H}H_L-\dot{H_L}}} \nonumber \\
 &&\hspace{2.0 cm} + \frac{Y_{ij}}{a^2}\PR{k^2 \delta \phi + \dot{\phi}\PC{kB - 2 \mathcal{H} H_T - \dot{H_T}}}  \\
 &&\delta \PC{\nabla_{0} \nabla_{j} \phi} = \frac{Y_j}{a^2}\PR{-k \dot{\delta \phi} + k \mathcal{H} \delta \phi + \dot{\phi}\PC{kA + \mathcal{H}B}}.
\end{eqnarray}
\end{widetext}

\end{document}